# Low-complexity Non-coherent Signal Detection for Nano-Scale Molecular Communications

Bin Li[1], Mengwei Sun[1], Siyi Wang[2], Weisi Guo[3], Chenglin Zhao[1]

**Abstract**

Nano-scale molecular communication is a viable way of exchanging information between nano-machines. In this letter, a low-complexity and non-coherent signal detection technique is proposed to mitigate the inter-symbol-interference (ISI) and additive noise. In contrast to existing coherent detection methods of high complexity, the proposed non-coherent signal detector is more practical when the channel conditions are hard to acquire accurately or hidden from the receiver. The proposed scheme employs the concentration difference to detect the ISI corrupted signals and we demonstrate that it can suppress the ISI effectively. The concentration difference is a stable characteristic, irrespective of the diffusion channel conditions. In terms of complexity, by excluding matrix operations or likelihood calculations, the new detection scheme is particularly suitable for nano-scale molecular communication systems with a small energy budget or limited computation resource.

## I. Introduction

MOLECULAR communications transmit information through the diffusion of chemical molecules [1]–[3]. Inspired by nature, the diffusion channel has advantages in both transmit energy efficiency and propagation loss over conventional electromagnetic (EM) wave channels, e.g. the total energy hitting the receiver suffers a lower rate of distance dependent loss [4], [5]. Perhaps this is one of the reasons why molecular communication systems are abundant in nature, both at the nano- and at the macro-scale [6]–[9]. For macro-scale applications (e.g. moths and crustaceans [7]), the molecular diffusion is often assisted by ambient flows, significantly accelerating the diffusion process. Furthermore, the data carried by diffusion channels are delay-tolerant. For nano-scale applications, the distance is small and the time delay remains within a few micro-seconds. As for the area of nano-medicine, most existing nano-machines lack the wireless communication capability to work effectively towards a common goal, e.g., targeted drug delivery or surgery. The aforementioned example of applying efficient micro- or nano-scale communications to nano-devices creates a new paradigm to future medical applications [10].

[1]Bin Li, Mengwei Sun and Chenglin Zhao are with the School of Information and Communications Engineering, Beijing University of Posts and Telecommunications (BUPT), Beijing 100876, China (E-mail: stonebupt@gmail.com).

[2]Siyi Wang is with the Department of Electrical and Electronic Engineering, Xi'an Jiaotong-Liverpool University, Suzhou 215123, China (Email: siyi.wang@xjtlu.edu.cn).

[3]Weisi Guo is with the School of Engineering, The University of Warwick, Coventry, CV4 7AL, UK (E-mail: weisi.guo@warwick.ac.uk).

Recently, a simple practical testbed has been developed in [11], one that can send generic text information over a few meters of unbounded free-space. The implementation consists of a transmitter that emits a short pulse of molecules bearing the information data using binary concentration shift keying (B-CSK) modulation, which was firstly introduced by M. S. Kuran et. al. [12]. The channel is a random walk diffusion channel with the induced drift. The receiver consists of a chemical detector, which will sense the concentration of molecules at a specific sample time and demodulate the signal. Since the diffusion channel impulse response (CIR) has a long tail, molecules from the previous emissions will inevitably intermix with the response of the current pulse, resulting in the inter-symbol interference (ISI). In some cases, this can result in aggregated ISI, when the chemical sensor cannot absorb incoming molecules [13]. As ISI is the dominant source of errors, one of the major challenges is in designing signal processing methods to combat ISI. Recently, the maximum *a posteriori* (MAP) and minimum mean square error (MMSE) schemes have been proposed in [14]. By acquiring the channel response, a decision-feedback equaliser (DFE) is designed. Similarly, transmitter side chemical pulse shaping methods proposed in [15] also demand knowledge of channel parameters. Such coherent methods could mitigate the ISI, but requiring accurate channel response estimation.

In this paper, we propose a novel non-coherent detection scheme for low-complexity molecular communication receivers. In sharp contrast to existing coherent schemes that require the accurate channel state information (CSI), the proposed scheme is premised essentially on the difference of the integrated concentration between two adjacent bins. The difference in practice is a stable characteristic, irrespective of the diffusion channel conditions. Despite its simple implementation, i.e., excluding any matrix operations and likelihood calculations that are indispensable in coherent schemes, the non-coherent scheme may also eliminate the ISI effectively. Numerical simulations are used to validate our presented scheme. The results demonstrated that the proposed non-coherent scheme may even be comparable to the coherent MAP and MMSE algorithms in good operating conditions, but does so with a much lower computational burden and requiring no channel information, which is hence attractive to practical molecular communication system integration.

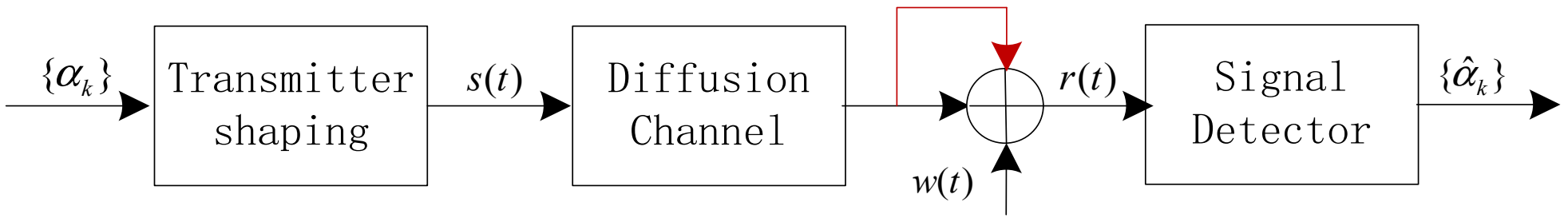

Fig. 1. A physical-layer schematic structure of molecules communication.

## II. System Model

In this section, we will first elaborate on the molecular signal modulation and channel model. This is then followed by a short introduction on some existing coherent detection schemes, e.g., MAP and MMSE.

### A. Single Model

A typical physical model of molecular communications is illustrated in Fig. 8. The nano-device transmitter emits a short pulse with the duration of $T_p$, with every $T_b$ interval. The on-off keying (OOK) is used along with a B-CSK modulation scheme [16], i.e., the emission is enabled if the binary symbol $\alpha_k \in \mathcal{A} = \{0, 1\}$ is 1, $k = 0, 1, \ldots, \infty$. Thus, the emitting pulse sequence is:

$$s(t) = \sum_{k=0}^{\infty} \alpha_k \times \text{rect}\left(\frac{t - T_p/2}{T_p} - kT_b\right). \tag{1}$$

Without loss of generality, in (1) a rectangular pulse shaper rect $(t)$ is adopted, i.e.,

$$\text{rect}\left(\frac{t - T_p/2}{T_p}\right) = \begin{cases} A, & -T_p/2 \leq t \leq T_p/2, \\ 0, & \text{otherwise}, \end{cases} \tag{2}$$

where $A$ represents the chemical concentration. After propagated via a diffusion channel $h(t)$ and contaminated by other imperfections, the received concentration is:

$$\begin{aligned} s(t) &= A \sum_{k=0}^{\infty} \alpha_k \cdot \text{rect}\left(\frac{t - T_p/2}{T_p} - kT_b\right) \otimes h(t) + w(t) \\ &= A \sum_{k=0}^{\infty} \alpha_k y(t - kT_b) + w(t), \end{aligned} \tag{3}$$

where the notation $\otimes$ denotes the convolution operation; $y(t - kT_b) = \text{rect}[(t - T_p/2)/T_p - kT_b] \otimes h(t)$ represents the equivalent channel response between the binary information source $\{\alpha_k\}$ and the receiver; $w(t)$ is the additive noise aroused by the imperfect counting process (when the low complexity chemical sensors with low sensitivity are considered), which is assumed to be an independent and identically distributed (i.i.d) zero-mean white Gaussian noise [17], with a variance of $\sigma_w^2$, i.e., $w(t) \sim \mathcal{N}(0, \sigma_w^2)$.

We consider a 3-D environment where the movements of molecules are characterised by the free diffusion (or Brownian motion) [18]. The well-established impulse response of such diffusion channels, by further considering the absorbing effect from a molecule receptor, will be given by [19], [20]:

$$h(t) = \frac{mR(d-R)}{d(4\pi t f)^{3/2}} \times \exp\left[-\frac{(d-R)^2}{4tf}\right], \tag{4}$$

where $m$ is the total number of emitted molecules; $R$ is the radius of the absorbing spherical receiver; $f$ is the diffusion coefficient of specific medium; $d$ is the Euclidean distance between transmitter and receiver [21]. As mentioned previously, unlike classical wireless channels, the diffusion CIR $y(t)$ will exhibit a very long trail given the specific $d$ and $R$.

For convenience, it is assumed the synchronisation of molecular communication links has been accomplished [22]. The nano-receiver will sample the molecules concentration inside the reception space at a Nyquist rate $R = 1/T_b$. The sampled discrete signal will be:

$$r_k = \sum_{m=0}^{\infty} \alpha_m y_{k-m} + w_k, \tag{5}$$

where $r_k = r(kT_b), y_{k-m} = y(kT_b - mT_b), w_k = w(kT_b)$. For $k - m < 0$, we will have $y_{k-m} = 0$. If the infinite ISI is truncated, the received signal may be expressed as:

$$r_k = \alpha_k y_0 + \sum_{m=k-I}^{k-1} \alpha_m y_{k-1-m} + w_k, \tag{6}$$

where the second term accounts for the ISI coming from the slow decay of previous CIRs. In [14], the ISI is combined with the additive Gaussian noise, which are then formulated as the non-stationary and signal dependent noise. Here, $I$ is defined as the truncating memory length, which specifies the interfering span of ISI from the previous transmitted pulses. Note that, whilst the value of $I$ is unbounded in practice, a finite value is used in this work for simulation and realistic algorithm purposes. For example, in [14], a finite memory is assumed in the sequential MAP and MMSE schemes.

### *B. Coherent Detectors*

*1) **MAP**:* The received vector is conveniently expressed as $\mathbf{r} = \mathbf{Ya} + \mathbf{w}$, where $\mathbf{r}_{K\times 1} \triangleq [r_k, r_{k-1}, \ldots, r_{k-K+1}]^T$; $\mathbf{a}_{(K+I)\times 1} \triangleq [a_k, a_{k-1}, \ldots, a_{k-K-I+1}]^T$; $\mathbf{Y}_{K\times(K+I)}$ is the *circulant* channel matrix constructed from $\mathbf{y}_{I\times 1}$, and $\mathbf{w}_{K\times 1}$ is additive noise vector. The channel matrix $\mathbf{Y}_{K\times(K+I)}$ is then expressed as:

$$\mathbf{Y} = \begin{bmatrix} y_0 & y_1 & \cdots & y_I & 0 & \cdots & 0 \\ 0 & y_0 & y_1 & \cdots & y_I & 0 & 0 \\ \vdots & \vdots & \vdots & \ddots & \vdots & \vdots & \vdots \\ 0 & \cdots & 0 & y_0 & y_1 & \cdots & y_I \end{bmatrix}.$$

For a coherent MAP scheme, the accurate estimation of unknown CSI, i.e. $\mathbf{y}_{I\times 1}$, should be acquired firstly. The estimation of unknown symbols, relying on the posterior density, is derived from:

$$\begin{aligned} \hat{\mathbf{a}}_{\text{MAP}} &= \arg\max_{\mathbf{a}\in\mathcal{A}^{K+I}} p(\mathbf{a}|\mathbf{r},\mathbf{Y}), \\ &= \arg\max_{\mathbf{a}\in\mathcal{A}^{K+I}} \prod_{k=1}^{K+I} p(\alpha_k|\alpha_{1:k-1}) \prod_{k=1}^{K+I} p(r_k|r_{1:k-1},\alpha_{1:k}), \end{aligned} \tag{7}$$

where the likelihood density $p(r_k|r_{1:k-1},\alpha_{1:k})$ follows a Gaussian distribution [14]. To reduce the complexity, a sequential estimation can be used in practice [23]. For the i.i.d information source $\alpha_k \in \mathcal{A}$ with the equal priors for “0” and “1”, the MAP scheme is equivalent to a maximum likelihood (ML) method.

*2)* ***MMSE:*** Another widely used sub-optimal detector is based on the MMSE criterion [23], which minimises the covariance matrix of estimation errors, i.e.,

$$\hat{\mathbf{a}}_{\text{MMSE}} = \arg\min_{\mathbf{a}\in\mathcal{A}^{K+I}} \mathbb{E}\left[(\mathbf{a}-\hat{\mathbf{a}})(\mathbf{a}-\hat{\mathbf{a}})^T\right]. \tag{8}$$

Based on the linear Gaussian model in (6), the MMSE estimation of the unknown information is derived via:

$$\begin{aligned} \hat{\mathbf{a}}_{\text{MMSE}} &= \mathbb{E}(\mathbf{a}|\mathbf{r}), \\ &= \mathbb{E}(\mathbf{a}) + (\mathbf{\Lambda}_w^{-1} + \mathbf{Y}^T\mathbf{\Lambda}_w^{-1}\mathbf{Y})^{-1}\mathbf{Y}^T\mathbf{\Lambda}_w^{-1}[\mathbf{r}-\mathbf{Y}\mathbb{E}(\mathbf{a})], \end{aligned} \tag{9}$$

where $\mathbb{E}(.)$ represents the statistical expectation; $\mathbf{\Lambda}_w$ is an $K\times K$ diagonal matrix with the elements of its principal diagonal are all $\sigma_w^2$. Note from (9) that, an accurate CIR is also required by the MMSE scheme.

## III. Non-coherent and Blind Signal Detections

It is recognised that, in sharp contrast to wireless communications, there are two inherent challenges in molecular communications. Firstly, the CSI cannot be easily estimated without a pilot channel, especially in the presence of channel disturbances due to air flow and temperature variations. Even in the presence of a pilot channel, the coherence period of the channel is usually small, whilst the channel delay is large, making the precise estimation a rough work. Secondly, molecular communication is targeted towards nano-scale systems, which naturally demand the low-computational complexity architecture to reduce hardware costs and energy expenditure. The effectiveness of the coherent MAP or MMSE algorithms in eliminating ISI largely depends on both accurate CSI and complicated matrix or polynomial (likelihood) operations. Based on the above considerations, those coherent detectors will be less attractive in the context of low-complexity and low-power nano-machines.

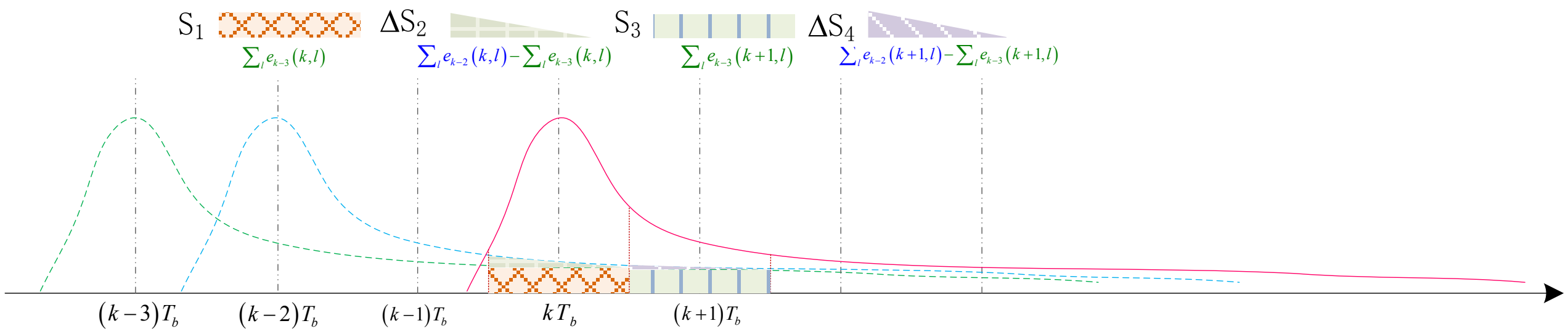

Fig. 2. The illustration of the proposed non-coherent detection schemes. Note that, each region will involve $2L+1$ samples.

### *A. Concentration Difference Estimator*

As far as we are aware of, for the first time a low-complexity non-coherent receiver is proposed for molecular communications by this work. The designed scheme essentially involve 3 steps.

*1)* ***Pre-smoothing****:* In order to suppress the additive noise, firstly a pre-smoothing (or pre-filtering) process will be used. To do so, the received molecular concentration is *over-sampled* around each symbol concentration (e.g., $kT_b$) with a higher rate $1/T_s$, as shown by Fig. 2. Then, a moving average process with a length of $2L+1$ is applied. Here, we assume $LT_s = T_b/2$. The smoothed signal $\widehat{r}_{k,l}$ is given by:

$$\widehat{r}_{k,l} = \sum_{l_0=-L}^{L} \frac{1}{2L+1} \times \tilde{r}_{k,l+l_0}. \tag{10}$$

Here, $\tilde{r}_{k,l}$ denotes the $l$th sub-sample around the $k$th symbol ($-L \leq l \leq L$). In practice, $L$ zeros will firstly be padded to both the front and the end of received signal sequence, i.e., $\tilde{\mathbf{r}} \triangleq [0, \ldots, 0, r_{1,-L}, r_{1,-L+1}, \ldots, r_{1,L}, \ldots, r_{K,-L}, r_{K,-L+1}, \ldots, r_{K,L}, 0, \ldots, 0]^T$.

*2)* ***Concentration Difference****:* In the new non-coherent scheme, the difference between concentration integrations is proposed to demodulate information, which is defined as:

$$\gamma_k \triangleq \sum_{l=-L}^{L} \widehat{r}_{k,l} - \sum_{l=-L}^{L} \widehat{r}_{k-1,l}. \tag{11}$$

In the following, we will show that, for a typical transmission rate $T_b = Nt_m$ (e.g. $N \geqslant 2$) where $t_m \propto d^2$ denotes the interval between the maximum concentration from the beginning of CIR, the suggested concentration difference (i.e. $\gamma_k$) may effectively suppress the ISI and achieve the promising detection performance, even if it can not mitigate it completely.

The molecules residue concentration at the time $k$, generated from the $(k-k_0)^{\text{th}}$ previous transmission, will be approximated by:

$$\begin{aligned} e_{k-k_0}(k,l) &= \alpha_{k-k_0} \times y[t-(k-k_0)T_b + LT_s] + b_{k-k_0} \times (L+l) + o\{(L+l)^2\}, \\ &\backsimeq C_{k-k_0} + b_{k-k_0} \times (L+l), \ \ -L \leqslant l \leq L. \end{aligned} \tag{12}$$

It is noted that, from the expression of channel response in eq. (4) and the illustration in Fig. 9, the ISI falling into subsequent time is not linearly varied with the index $k$, leading to great difficulties in the algorithm designing and analysis. From Eq. (12), it is shown that, fortunately, the ISI from previous molecular concentrations could be approximated by the first-order Taylor series expansion (TSE). Although it may underestimate ISI by dropping the high-order series, for the channel response $y(n)$ with a long trail, the above approximation is sufficient for the practice algorithm developments and analysis. With the above Eq. (12), now the nonlinear interference curve will be replaced by a *linear* function of a negative slope, and the ISI from the $(k-k_0)$th time is then approximated via the area under this line, i.e., $ISI \simeq \sum_{l=-L}^{L} e_{k-k_0}(k,l) \simeq \Delta S_1 + \Delta S_2$ with $a_{k-k_0} \in \{0,1\}$. Corresponding to the first and second terms in eq. (12), the ISI (i.e. a trapezoid area ) can be broken into a square area $\Delta S_1$ and the other triangle area $\Delta S_2$, as in Fig. 9. It is easy to see the constant $C_{k-k_0}$ in eq. (12) accounts for the *intercept* (or the width of a shadow rectangular $S_1$), while $b_{k-k_0} \simeq \alpha_{k-k_0} \times \frac{dy(t)}{dt}\Big|_{t=(k-k_0)T_b+LT_s}$ in the second term denotes the linear *slope*.

Based on Eq. (12), during the $k$th time bin, the accumulated concentrations can be further written to:

$$\begin{aligned} \sum_{l=-L}^{L} \widehat{r}_{k,l} &= \sum_{l=-L}^{L} \left[ \alpha_k y(t-kT_b-LT_s) + \sum_{k_0=1}^{I} e_{k-k_0}(k,l) \right], \\ &\approx \sum_{l=-L}^{L} \alpha_k y(t-kT_b-LT_s) + 2L\sum_{k_0=1}^{I} C_{k-k_0} + 2\sum_{k_0=1}^{I} b_{k-k_0} L^2, \\ &= \underbrace{\sum_{l=-L}^{L} \alpha_k y(t-kT_b-LT_s)}_{\text{useful signal concentration}} + \underbrace{2L\sum_{k_0=1}^{I} C_{k-k_0}}_{\text{ISI concentration}} + o(IL^2). \end{aligned} \tag{13}$$

Here, we omit the additive noise for convenience. The third term may be dropped as the slope is usually very small, i.e. $b_{k-k_0} \to 0$ when the $k_0$ is large. The concentration difference $\gamma_{k+1}$ is now reformatted by:

$$\begin{aligned} \gamma_{k+1} &= \sum_{l=-L}^{L} \widehat{r}_{k+1,l} - \sum_{l=-L}^{L} \widehat{r}_{k,l}, \\ &\overset{(a)}{\approx} \sum_{l=-L}^{L} \alpha_{k+1} y\left[t-(k+1)T_b - lT_s\right] - \sum_{l=-L}^{L} \alpha_k y(t-kT_b-lT_s), \end{aligned} \tag{14}$$

where (a) follows the fact that:

$$2L\sum_{k_0=1}^{I} C_{k+1-k_0} - 2L\sum_{k_0=1}^{I} C_{k-k_0} \to 0. \tag{15}$$

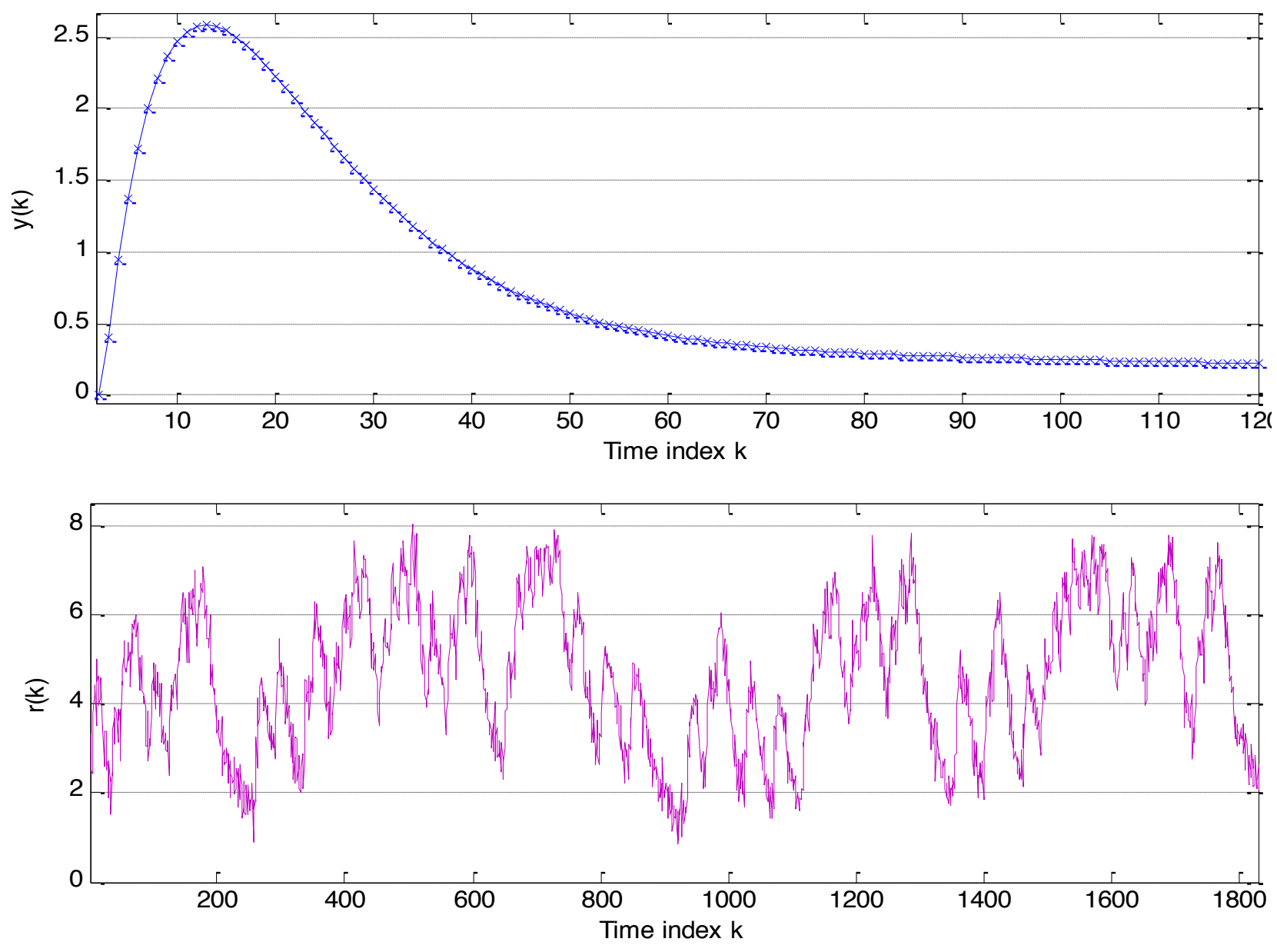

Fig. 3. The illustration of an instantaneous realisation of a molecular diffusion channel $y(t)$, as seen the top sub-figure, and an example of received signals $r(t)$ in the below sub-figure.

From the relationship of eq. (14), it is noted that the suggested metric $\gamma_{k+1}$ involves only the differences of the useful signals. In other words, the additive ISI has been eliminated basically. A physical explanation to eq. (14) is that, as suggested from Fig. 9, the ISI integration fallen into the time bin $[kT_b - LT_s, kT_b + LT_s]$ from the $(k-k_0)^{\text{th}}$ molecules emission will be approximated by the combined area of the trapezoid, i.e., $S_1 + \Delta S_2$. Similarly, the ISI integration fallen to $[(k+1) \times T_b - LT_s, (k+1) \times T_b + LT_s]$ may be approached via $S_3 + \Delta S_4$. It is found that the rectangular area $S_1$ is almost equal to $S_2$ and, therefore, we may have $S_1 + \Delta S_2 - (S_3 + \Delta S_4) \simeq S_1 - S_2 \simeq 0$.

*3) **Post-smoothing**:* A post-smoothing process will be used to further reduce additive noise, and a similar moving average process may be utilised. Finally, a threshold will be specified to determine whether there contains sufficient energy in the current time index $k$, indicating whether the new molecule concentration appears. In practice, the threshold is simply set to 0.

### *B. Complexity Analysis*

From the above elaborations, it is seen that the summation operations of our non-coherent scheme is about $\mathcal{O}(L)$ in complexity, i.e., requiring *no* multiplication. For the traditional MAP scheme, the numbers of multiplication are usually measured by $\mathcal{O}(2^I\theta)$, where $\theta$ accounts for the complexity of evaluating the likelihood function [24]. For the linear MMSE scheme, the complexity is $\mathcal{O}(I^3)$. For the typical configuration of $L = 5$ and $I > 20$, we may easily note that the proposed non-coherent scheme superior in its low implementation complexity.

## IV. Numerical Experiments

In the simulation setup, the maximum concentration of the channel response $y(t)$ is located at $t_m = 14T_s$[1], as seen from the Fig. 10-(a). The duration between two binary symbols is $T_b = 20T_s$. As shown in Fig. 10-(b), after diffused propagation, serious ISI will occur at the receiver. In the analysis, we consider ISI with a relatively long interfering length of $I = 30$. The moving average length is set to $L = \frac{T_b}{2T_s}$. Each bit error ratio (BER) curve is derived numerically based on $10^6$ symbols. Similar to [14], we treat ISI as the receiving self-interference. Given a total length of information bits $K_N$, the SNR in the presence of additive counting noise is alternatively defined as:

$$\text{SNR} \triangleq 10\log_{10}\left[\mathbb{E}\left(\sum_{k=1}^{K_N} r_k^2\right)/K_N\sigma_w^2\right]. \tag{16}$$

As suggested by [14], there also contains the ISI in the received signal $r(k)$. Another metric, known as the signal to noise and interference ratio (SINR), will be also investigated, as given by:

$$\text{SINR} \triangleq 10\log_{10}\left\{\frac{\mathbb{E}\left(\sum\limits_{k=1}^{K_N}\sum\limits_{l=-L}^{L} y(t-kT_b-lT_s)^2\right)}{\mathbb{E}\left\{\sum\limits_{k=1}^{K_N}\sum\limits_{l=-L}^{L}[r(t-kT_b-lT_s)-y(t-kT_b-lT_s)]^2\right\}}\right\}. \tag{17}$$

### A. Performance Analysis

From the Fig. 11, we may observe that, given a relatively high data rate, e.g. $T_b = 20T_s = 1.43t_m$, the non-coherent algorithm will obtain the promising detection performance. It is found that, by suppressing the ISI to the minimum, the sequential MAP scheme will achieve the best performance in the high SNR regime. Although the proposed non-coherent detector has a very low-complexity implementation, the new scheme is able to achieve a superior performance in comparison with the coherent MMSE scheme, but cannot achieve a better performance than the coherent MAP scheme. More importantly, as a convenience the CSI can be excluded from our *blindly* non-coherent detection scheme, unlike classical coherent MAP or MMSE schemes.

In practice, the symbol duration $T_b$ may have significant influences on the detection performance. Firstly, we note from Fig. 9 that, in the low SNRs region (e.g., <16 dB with $T_b = 20T_s$), the non-coherent scheme will surpass the two coherent schemes. In high SNRs region, however the coherent MAP scheme will be superior to the non-coherent detector. Primarily this is because the designed non-coherent scheme is relatively robust to additive noises (e.g. with a moving average process), while the coherent schemes are more sensitive to such contaminations. Secondly, it is observed that, as the symbol duration $T_b$ decreases (or the transmission rate increases), the performance of all three methods will be degraded. Taking $L = 7$ ($T_b = 14T_s = t_m$) as an example, the detection performance of the non-coherent scheme will further be worse than both MAP and MMSE techniques. This is mainly attributed to the fact that, with a small $L$, the ISI of two successive time-bins (especially when $k_0 = 1$) could not be approximated by its first-order TSE anymore. In such a case, the concentration difference may not mitigate the ISI as effectively as set out in Eq. (14), leading to the serious performance deterioration. For coherent MAP and MMSE schemes, the residual errors will also be more apparent in the presence of more severe ISI.

[1]Note that, here we only focus on the processing of discrete samples. While the sampling time $T_s$ may depend on different applications, the signal processing scheme is independent of any specific $T_s$.

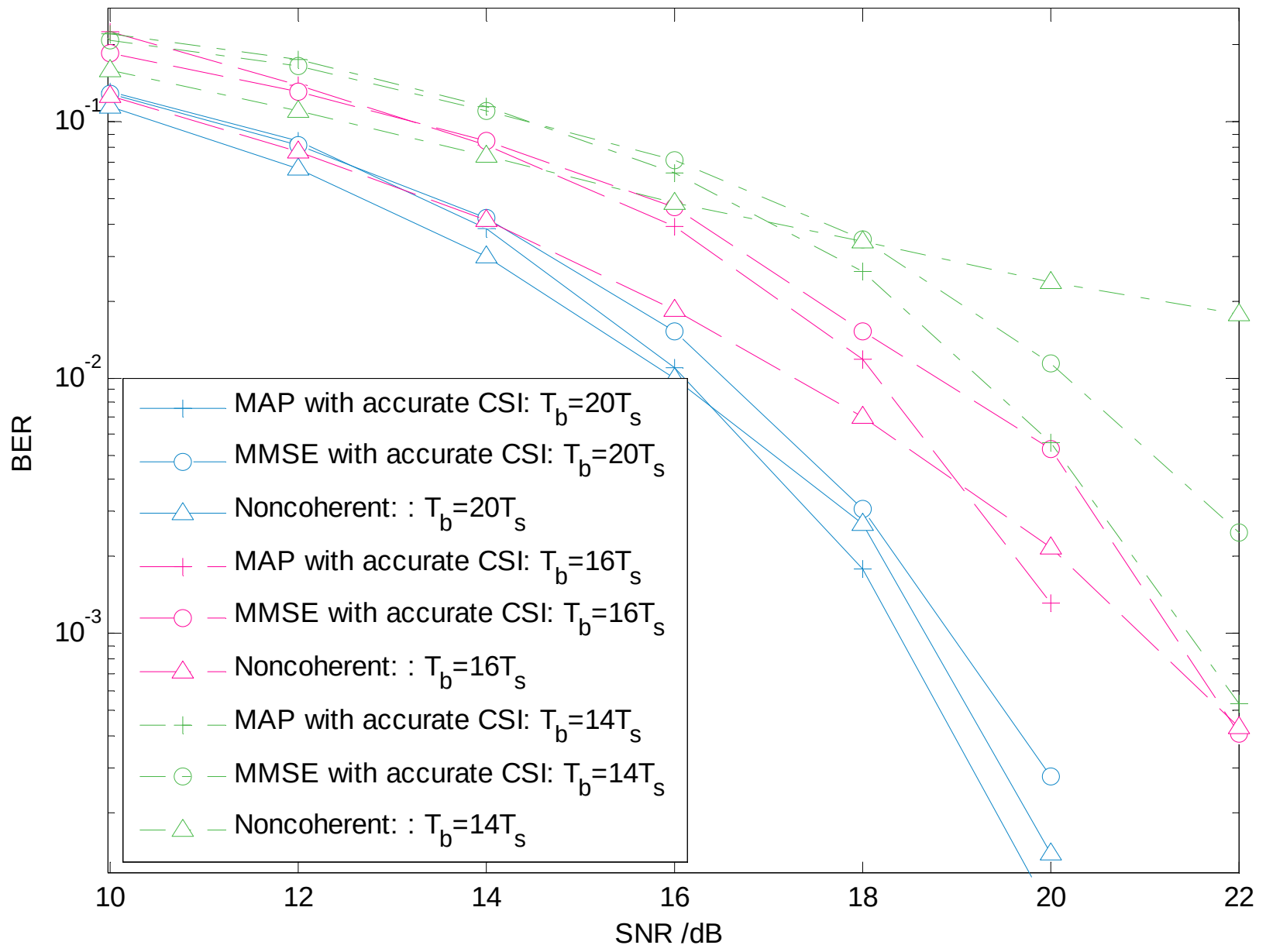

Fig. 4. BER performance of coherent detection schemes (MAP and MMSE) and the non-coherent scheme symbol interval. Note that, we assume $L = \frac{T_b}{2T_s}$.

For some specific molecular applications calling for the low implementation complexity and low operational power expenditure, the memory length of ISI in algorithm processing should be restrained when implementing MAP or MMSE algorithms. If the processing ISI length is restricted to $I = 10$, from Fig. 12 the proposed non-coherent scheme will surpass both MAP and MMSE detectors, resulting in a superior ISI mitigation performance.

Furthermore, as the ISI is treated as the receiving interference, we may investigate the BER performance of different detection schemes in the context of SINRs. As seen from Fig. 10, the channel response $y(n)$ is assumed to exhibit a long trail, the self-interference from previous transmissions, therefore, will dominate the molecule receiver. In addition, we have considered another Gaussian noise to characterise the sensing (or counting) errors of a chemical receptor. It is found from Fig. 13 that, given a long-trail ISI response (e.g. $I = 30$) and the equal probability of 1s and 0s in an OOK line-code, the SINR will converge to a constant value (e.g. $-3$ dB for the considered symbol period) as the additive error tends to be 0, indicating the ISI will always surpass useful signals, which may remain quite different in comparison to wireless communications. From the simulation result, we observed that, in realistic molecular communications, the proposed non-coherent detection scheme will become comparable to both MAP and MMSE as the SINR approaching $-3$ dB (or the variance of additive noises approaching 0).

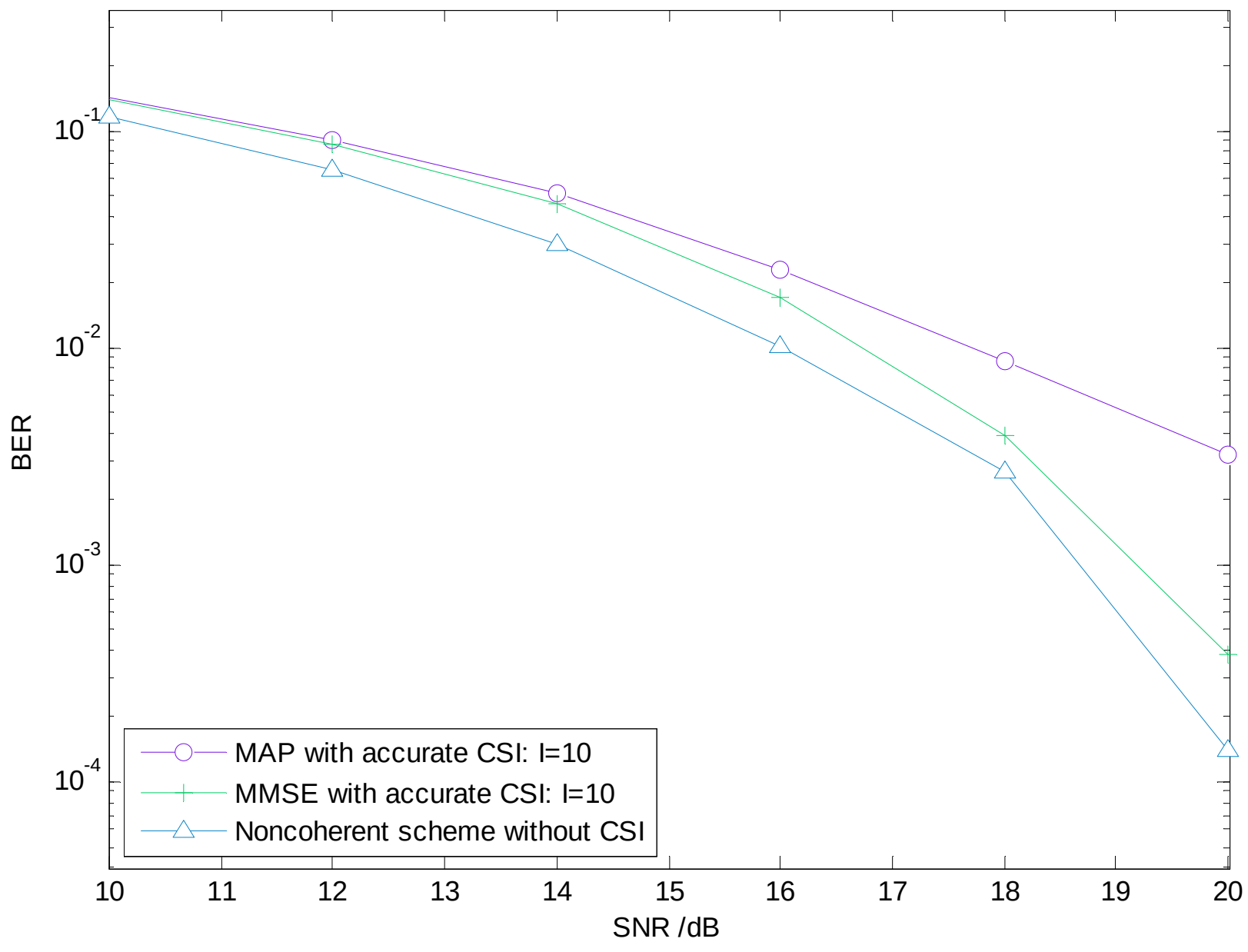

Fig. 5. BER performance of coherent detection schemes (MAP and MMSE) and the noncoherent scheme under various $I$.

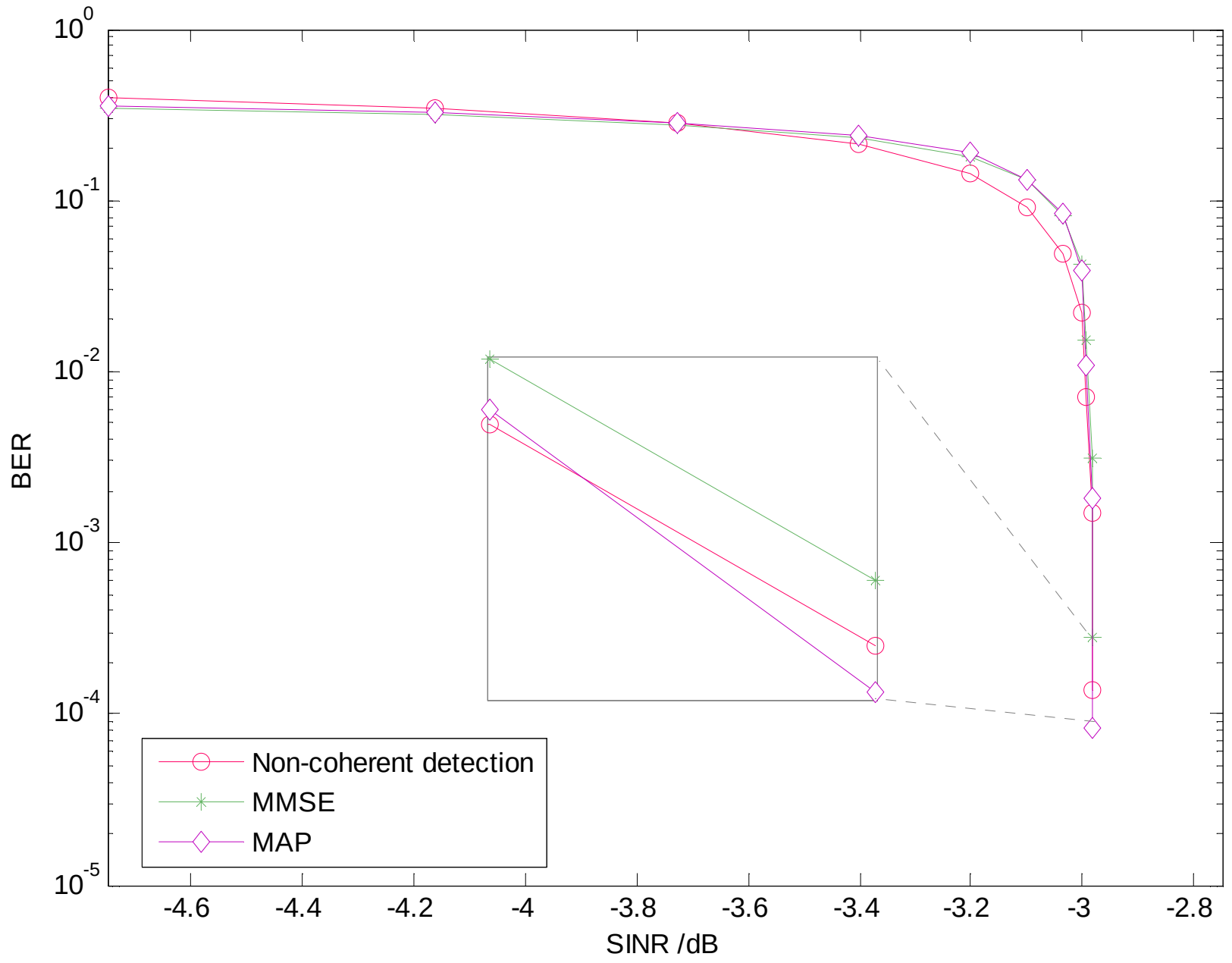

Fig. 6. BER performance of various detection schemes under different SINR.

### *B. Comparative Analysis*

Note that, the modulation scheme investigated in the above analysis can be viewed as a special case of the binary-concentration shift keying (B-CSK) [12], i.e., no molecule will be transmitted in the case of “0”. As suggested by ref. [12], a one-step sequential interference cancellation (SIC) scheme would be adopted to mitigate ISI from the last time-slot. On this basis, the information capacity has also been derived based on a simplified channel model, provided the interference to the time $k$ from previous slots $k-k_0$ ($k_0 \geq 2$) could be ignored. These previous works in [12], [25] may be of great importance to the low data-rate molecular communications (i.e., the symbol duration is greatly larger than $t_m$). As far as some scenarios of relatively high data-rate are concerned, however, the ISI from multiple previous slots will occur at time $k$, as seen from Fig. 10, which would make the one-step SIC detection infeasible. Given the ISI channel of $I = 30$, the numerically derived BER curve are plotted in Fig. 14. It is seen that, for one thing, the one-step SIC scheme fails to deal with a long ISI channel, and the BER value can only decrease to a floor of 7.4 $\times 10^{-2}$. For another, even if the SIC method is generalised to multiple steps, the BER performance will be converged to a floor of 6.3 $\times 10^{-4}$ if the SIC span surpasses 5. It is considered that the BER floors are mainly aroused by the *error propagation* of SIC. That is, once a detection error occurs, it will affect the detection of subsequent time slots. Besides, the threshold configuration after multiple-steps SIC requires full information of CIR, making it less attractive in low-complexity applications. From the numerical results, the proposed non-coherent detection scheme will be inferior to the multiple-steps SIC in low SNRs. By mitigating the BER floor effectively, however the new non-coherent scheme will be more competitive in high SNRs.

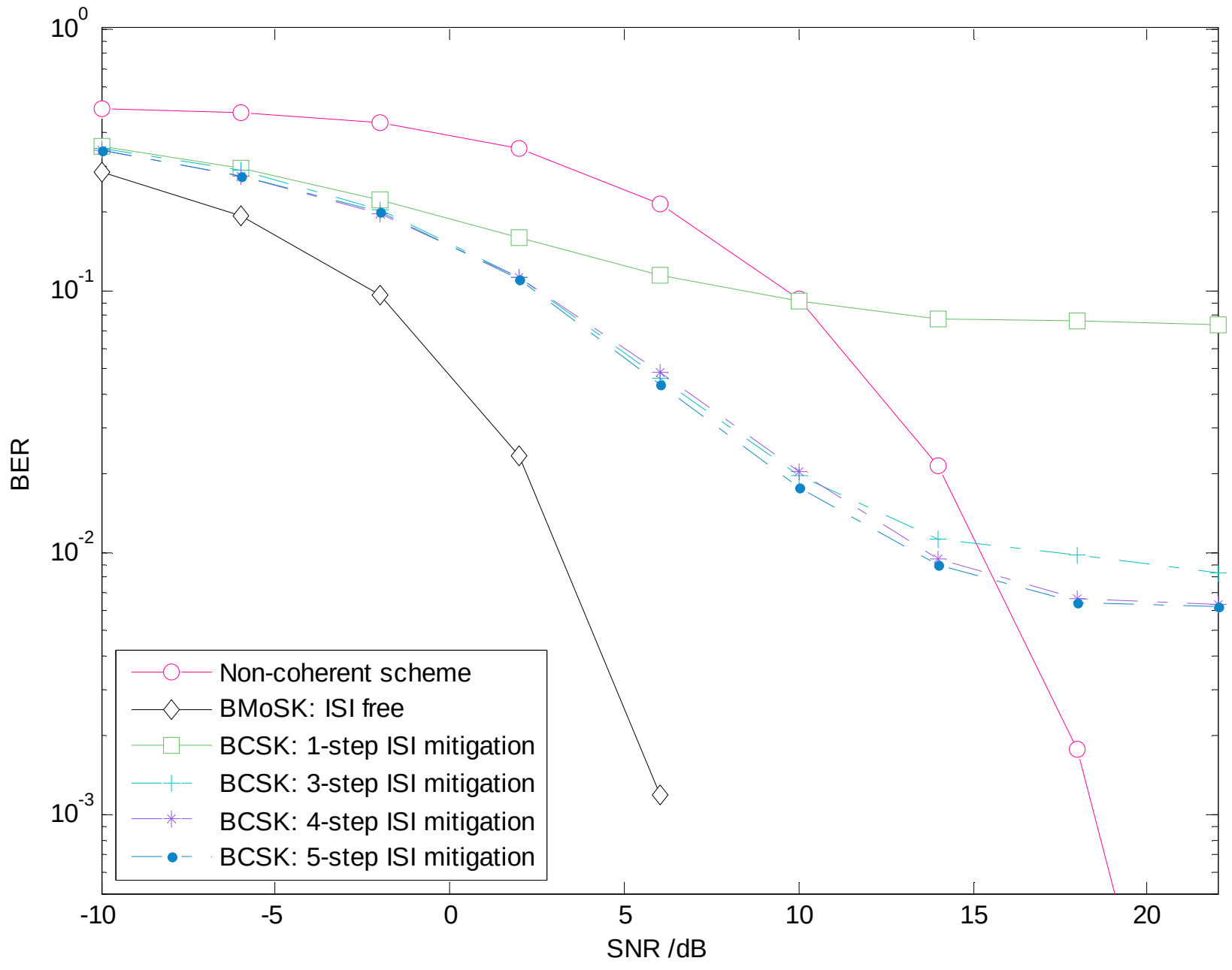

Fig. 7. BER performance of different modulation schemes.

For the comprehensive analysis, we further derive numerically the BER performance of another modulation scheme, i.e., binary molecule shift keying (B-MoSK) [12], [25], in which two distinct molecules are utilised to convey two different information bits (or messages). It is seen that the interference from the previous time slot (i.e. $k-1$) would be fully mitigated under a simplified channel model (i.e. with the 1st-order ISI). For the ideal case, the B-MoSK modulation scheme will achieve the promising detection performance by mitigating ISI via modulation techniques, as demonstrated by Fig. 14. When the long ISI channel is considered (i.e. corresponding to a small symbol duration), however the interference from previous time $(k-1-k_0)$ $(k_0>1)$ may appear at the current time $k$. Even if a similar SIC technique can be applied, the BER performance of the B-MoSK modulation will be degraded then. Moreover, with the integrated two different chemical receptors, the implementation of a B-MoSK receiver may easily be impractical to certain low-complexity scenarios. Meanwhile, note that if these two molecule receptors are not completely separable, then the adjacent information corruption will further occur. In contrast, the non-coherent scheme, which allows a compromise between the complexity and performance, will be of great promise to the low-complexity and robust molecular communications.

## V. Conclusions

A low-complexity and non-coherent signal detector is proposed to mitigate ISI in the diffusion-based nano-scale molecular communication systems. This new scheme is premised on the difference of molecules concentrations, which is a stable characteristic irrespective of channel conditions. Therefore, this method excludes the need for accurate and timely channel estimations. Both the theoretical analysis and experimental simulations demonstrate that, provided a moderate transmission rate (e.g., $T_b = 1.4t_m$) and a long ISI channel, the proposed non-coherent scheme is able to achieve the comparable error rate performance compared with coherent MMSE or MAP schemes. The proposed non-coherent detector is of great promise to low-complexity and low-rate molecular communications, as it completely excludes the matrix operations and likelihood calculations and thereby has the low computational and implementation complexity. Future work includes its extensions to extremely high rates ($T_b \leq t_m$) and the hardware implementation.

## Acknowledgment

This work of B. Li, M. Sun and C. Zhao was supported by Natural Science Foundation of China (NSFC) under Grants 61471061 and the Fundamental Research Funds for the Central Universities under Grant 2014RC0101. The work of S. Wang was in part supported by the Research Development Fund (RDF-14-01-29) of Xi'an Jiaotong-Liverpool University.

## References

[1] S. Hiyama, Y. Moritani, T. Suda, R. Egashira, A. Enomoto, M. Moore, and T. Nakano, "Molecular communication," in *Proceedings of the NSTI Nanotechnology Conference and Trade Show*, 2005, pp. 391–394.

[2] L. Parcerisa Giné and I. F. Akyildiz, "Molecular communication options for long range nanonetworks," *Computer Networks*, vol. 53, no. 16, pp. 2753–2766, 2009.

[3] T. Nakano, A. Eckford, and T. Haraguchi, *Molecular Communications*. Cambridge University Press, 2013.

[4] I. Llaster, A. Cabellos-Aparicio, M. Pierobon, and E. Alarcon, “Detection Techniques for Diffusion-based Molecular Communication,” *IEEE Journal on Selected Areas in Communications (JSAC)*, vol. 31, no. 12, pp. 726–734, Jan. 2014.

[5] W. Guo, C. Mias, N. Farsad, and J.-L. Wu, “Molecular versus electromagnetic wave propagation loss in macro-scale environments,” *IEEE Transactions on Molecular, Biological and Multiscale Communications*, 2015.

[6] A. Mafra-Neto and R. Carde, “Fine-scale structure of pheromone plumes modulates upwind orientation of flying moths,” *Nature*, vol. 369, pp. 142–144, 1994.

[7] T. Wyatt, *Pheromones and Animal Behaviour: Communication by Smell and Taste*. Cambridge University Press, 2003.

[8] T. Nakano, M. J. Moore, F. Wei, A. V. Vasilakos, and J. Shuai, “Molecular communication and networking: Opportunities and challenges,” *IEEE Transactions on NanoBioscience*, vol. 11, no. 2, pp. 135–148, 2012.

[9] T. Nakano, T. Suda, Y. Okaie, M. J. Moore, and A. V. Vasilakos, “Molecular communication among biological nanomachines: A layered architecture and research issues,” *IEEE Transactions on NanoBioscience*, vol. 13, no. 3, pp. 169–197, 2014.

[10] B. Atakan, O. B. Akan, and S. Balasubramaniam, “Body area nanonetworks with molecular communications in nanomedicine,” *IEEE Communications Magazine*, vol. 50, no. 1, pp. 28–34, 2012.

[11] N. Farsad, W. Guo, and A. W. Eckford, “Tabletop molecular communication: Text messages through chemical signals,” *PloS one*, vol. 8, no. 12, p. e82935, 2013.

[12] M. S. Kuran, H. B. Yilmaz, T. Tugcu, and I. F. Akyildiz, “Modulation techniques for communication via diffusion in nanonetworks,” in *2011 IEEE International Conference on Communications (ICC)*. IEEE, 2011, pp. 1–5.

[13] S. Wang, W. Guo, S. Qiu, and M. D. McDonnell, “Performance of Macro-Scale Molecular Communications with Sensor Cleanse Time,” in *International Conference on Telecommunications (ICT)*, May 2014, pp. 363–368.

[14] D. Kilinc and O. B. Akan, “Receiver Design for Molecular Communication,” *IEEE Journal on Selected Areas in Communications*, vol. 31, no. 12, pp. 705–714, 2013.

[15] S. Wang, W. Guo, and M. D. McDonnell, “Transmit pulse shaping for molecular communications,” in *Proceedings IEEE INFOCOM*, 2014, pp. 209–210.

[16] M. Pierobon and I. F. Akyildiz, “Information capacity of diffusion-based molecular communication in nanonetworks,” in *Proceedings IEEE INFOCOM*, 2011, pp. 506–510.

[17] N. Farsad, N. Kim, A. Eckford, and C. Chae, “Channel and noise models for nonlinear molecular communication systems,” *IEEE Journal on Selected Areas in Communications (JSAC)*, pp. 1–14, Nov. 2014.

[18] B. Oksendal, *Stochastic Differential Equations: An Introduction with Applications*. Springer, 2010.

[19] H. B. Yilmaz, A. C. Heren, T. Tugcu, and C.-B. Chae, “Three-Dimensional Channel Characteristics for Molecular Communications With an Absorbing Receiver,” *IEEE Communications Letters*, vol. 18, no. 6, pp. 929–932, 2014.

[20] M. S. Leeson and M. D. Higgins, “Forward error correction for molecular communications,” *Nano Communication Networks*, vol. 3, no. 3, pp. 161–167, 2012.

[21] M. Pierobon and I. F. Akyildiz, “A physical end-to-end model for molecular communication in nanonetworks,” *IEEE Journal on Selected Areas in Communications*, vol. 28, no. 4, pp. 602–611, 2010.

[22] H. ShahMohammadian, G. Messier, and S. Magierowski, “Blind synchronization in diffusion-based molecular communication channels,” *IEEE communications letters*, vol. 17, no. 11, pp. 2156–2159, 2013.

[23] S. Kay, *Fundamentals of statistical signal processing, volume 1: Estimation theory*. Prentice Hall, 1998.

[24] B. Haible and T. Papanikolaou, “Fast multiprecision evaluation of series of rational numbers,” in *Proceedings of the Third International Symposium on Algorithmic Number Theory (ANTS)*. Springer, Jun. 1998, pp. 338–350.

[25] H. Arjmandi, A. Gohari, M. N. Kenari, and F. Bateni, “Diffusion-based nanonetworking: A new modulation technique and performance analysis,” *IEEE Communications Letters*, vol. 17, no. 4, pp. 645–648, 2013.